\documentclass[twocolumn,superscriptaddress,showpacs,pra]{revtex4}
\usepackage{epsfig}
\usepackage{delarray}
\usepackage{amsmath}
\usepackage{bm}

\newcommand{\braket}[2]{\langle #1 \,|\, #2 \rangle}
\newcommand{\ket}[1]{| \, #1 \rangle}
\newcommand{\bra}[1]{ \langle #1 \,  |}

\begin{document}
\title{Probabilistic cloning with supplementary information}

\author{Koji Azuma}
\email{azuma@qi.mp.es.osaka-u.ac.jp}
\affiliation{Division of Materials Physics, Department of Materials Engineering Science, 
Graduate School of Engineering Science, Osaka University, Toyonaka, Osaka 560-8531, Japan}

\author{Junichi Shimamura}
%\email{} 
\affiliation{Division of Materials Physics, Department of Materials Engineering Science, Graduate School of Engineering Science, Osaka University, Toyonaka, Osaka 560-8531, Japan}
\affiliation{CREST Photonic Quantum Information Project, 4-1-8 Honmachi, Kawaguchi, Saitama 331-0012, Japan}
\affiliation{SORST Research Team for Interacting Carrier Electronics, 4-1-8 Honmachi, Kawaguchi, Saitama 331-0012, Japan}

\author{Masato Koashi}
%\email{}
\affiliation{Division of Materials Physics, Department of Materials Engineering Science, Graduate School of Engineering Science, Osaka University, Toyonaka, Osaka 560-8531, Japan}
\affiliation{CREST Photonic Quantum Information Project,
4-1-8 Honmachi, Kawaguchi, Saitama 331-0012, Japan}
\affiliation{SORST Research Team for Interacting Carrier Electronics,
4-1-8 Honmachi, Kawaguchi, Saitama 331-0012, Japan}

\author{Nobuyuki Imoto}
%\email{}
\affiliation{Division of Materials Physics, Department of Materials Engineering Science, Graduate School of Engineering Science, Osaka University, Toyonaka, Osaka 560-8531, Japan}
\affiliation{CREST Photonic Quantum Information Project,
4-1-8 Honmachi, Kawaguchi, Saitama 331-0012, Japan}
\affiliation{SORST Research Team for Interacting Carrier Electronics,
4-1-8 Honmachi, Kawaguchi, Saitama 331-0012, Japan}

\date{\today}

\begin{abstract}
We consider probabilistic cloning of a state chosen
from a mutually nonorthogonal set of pure states,
with the help of a party holding 
supplementary information in the form of pure states.
When the number of states is 2, we show that 
the best efficiency of producing $m$ copies is always achieved 
by a two-step protocol in which the helping party 
 first attempts to produce $m-1$ copies from the supplementary state, 
and if it fails, then the original state is used to produce $m$ copies.
On the other hand, when the number of states exceeds two,
the best efficiency is not always achieved by such a protocol.
We give examples in which the best efficiency is not achieved 
even if we allow   
any amount of one-way classical 
communication from the helping party.
\end{abstract}

\maketitle

%%%%%%%%%%%%%%%%%%%%%%%%%%%%%%% main %%%%%%%%%%%%%%%%%%%%%%%%%%%%%

\section{Introduction}
The impossibility of deterministic cloning of nonorthogonal pure states
 is well known as the no-cloning theorem
\cite{WZ82,Y86}. The best one can do is to carry out weaker tasks,
such as allowing the copies to be inaccurate \cite{BH96,MPH97,GH97,BVOKH97,HB97,BDEMS98,GM97,BEM98}, or allowing 
a failure to occur with a nonzero probability (probabilistic cloning)
\cite{DG98}. Another way to enable the cloning is to provide 
some hints in the form of a quantum state. Jozsa has 
 considered \cite{J02}
how much or what kind of supplementary information $\hat{\rho_i}$ is
required to make two copies $\ket{\psi_i}\ket{\psi_i}$ from the original
information $\ket{\psi_i}$. He has shown that for any mutually
nonorthogonal 
set of 
original states $\{\ket{\psi_i} \}$, whenever two copies $\ket{\psi_i}\ket{\psi_i}$ are generated with the help of the supplementary information $\hat{\rho_i}$, the state $\ket{\psi_i}$ can be generated from the supplementary information $\hat{\rho_i}$ alone, independently of the original state, i.e.,
\begin{equation}
\ket{\psi_i} \otimes \hat{\rho_i} \stackrel{{\footnotesize \mbox{CPTP}}}{\longrightarrow} \ket{\psi_i}\ket{\psi_i} \Longrightarrow  \hat{\rho_i} \stackrel{{\footnotesize \mbox{CPTP}}}{\longrightarrow } \ket{\psi_i},
\end{equation}
where CPTP stands for a completely positive trace-preserving map,
implying that the transformation can be done deterministically.
This result, dubbed the stronger no-cloning theorem,
implies that the supplementary information must be provided in the form of the result $\ket{\psi_i}$ itself, rather than a help,
thereby obliterating the necessity 
of the cloning task itself.

An interesting question occurring here is whether we can find
a similar 
property in the case of probabilistic cloning when
we ask how much increase in the success probability is 
obtained with the help of supplementary information.
Suppose that the success probability of cloning
the $i$th state $\ket{\psi_i}$ without any help 
is $\gamma_i$. If we are directly given a right copy 
of state $\ket{\psi_i}$ with probability $q_i$, 
the success probability would increase to 
$\gamma'_i=q_i+(1-q_i)\gamma_i$. Hence 
the counterpart of the stronger no-cloning theorem  
in probabilistic cloning will be the implication 
\begin{multline}
\ket{\psi_i} \otimes \hat{\rho_i} \stackrel{\gamma'_i}{\longrightarrow} \ket{\psi_i}\ket{\psi_i} \\\Longrightarrow  \hat{\rho_i} \stackrel{q_i}{\longrightarrow } \ket{\psi_i},\; \ket{\psi_i}\stackrel{\gamma_i}{\longrightarrow } \ket{\psi_i}\ket{\psi_i}
\end{multline}
with $\gamma'_i=q_i+(1-q_i)\gamma_i$. 
In other words, it implies that the best usage of the supplementary 
information is to probabilistically create a copy $\ket{\psi_i}$
from it, independently of the original state.

If there are cases where the above implication is not true,
it follows that the supplementary information can help 
directly the process of the cloning task in those cases. 
Then, the next 
question will be to ask what kind of interaction should occur 
between the supplementary information and the 
original information. 

In this paper, we consider probabilistic cloning of mutually 
nonorthogonal pure states when 
supplementary information is given as a
pure state. We prove that 
when the number of the possible original states
is 2, the above implication is true, namely, the 
supplementary information only serves to provide a copy with 
a nonzero probability and it does not directly help the 
process of the cloning. On the other hand, when we
have more than two states to choose from, the above 
implication is not always true. To see this,
it is convenient to assume two parties, Alice and Bob, respectively
holding the original information and the supplementary information. 
We give examples in which there is a gap 
between the efficiency when Bob only communicates 
to Alice with a one-way classical channel and the
efficiency when they fully cooperate through a 
quantum channel.

This paper is organized as follows. In Sec. \ref{se:2},
we provide definitions and basic theorems used in later
sections. We discuss the two-state problem in Sec. \ref{se:3} and 
prove that the property similar to the stronger no-cloning
theorem holds in this case. In Sec. \ref{se:4},  
we give examples with three or more states and show that 
there is a gap between the success probabilities in the 
scenarios with classical communication and 
quantum communication. Section \ref{se:5} concludes the paper.

\section{Probabilistic Transformation Theorem}\label{se:2}

Throughout this paper, we consider a class of machines 
that conducts probabilistic transformation of input pure states
into output pure states. We denote by 
$\{\ket{\Phi_i}\stackrel{\gamma_i}{\longrightarrow}\ket{\Psi_i}\}_{i=1,\ldots,n}$
a machine having the following properties. (i) It receives
 a quantum state as an input, and returns a quantum state 
as an output, together with one bit of classical output
indicating whether the transformation has been successful or not.
(ii) When the input quantum state is $\ket{\Phi_i}$, the transformation
succeeds with probability $\gamma_i$, and the successful 
output state is $\ket{\Psi_i}$.
Note that if the output states $\{ \ket{\Psi_i} \}$ form an orthonormal
set,
namely,
$(\forall i,j)\left(\braket{\Psi_i}{\Psi_j}=\delta _{ij} \right)$, the 
machine carries out unambiguous discrimination of the set $\{
\ket{\Phi_i} \}$
with success probabilities $\{\gamma_i\}$.

A necessary and sufficient condition for the existence of
a machine 
$\{\ket{\Phi_i}\stackrel{\gamma_i}{\longrightarrow}\ket{\Psi_i}\}_{i=1,\ldots,n}$
is given by the following theorem.
%(Appendix \ref{ape1} gives
%a sketch of the proof). 

{\it Theorem 1.}
There exists a machine 
$\{\ket{\Phi_i}\stackrel{\gamma_i}{\longrightarrow}\ket{\Psi_i}\}_{i=1,\ldots,n}$
if and only if there are normalized states $ \ket{P^{(i)}} \;(i=1,\ldots,n)$ such that 
the matrix $X- \sqrt{\Gamma} Y \sqrt{\Gamma}$ is positive semidefinite, where $X:=[\braket{\Phi_i}{\Phi_j}]$, $Y:=[\braket{\Psi_i}{\Psi_j} \braket{P^{(i)}}{P^{(j)}}]$ and $\Gamma:={\rm diag}(\gamma_1,\gamma_2,\ldots,\gamma_n)$ are  $n \times n$ matrices.

This theorem can be proved by a similar way as in the probabilistic cloning
theorem by Duan and Guo \cite{DG98}.
A general description of a machine
$\{\ket{\Phi_i}_A\stackrel{\gamma_i}{\longrightarrow}\ket{\Psi_i}_A\}_{i=1\ldots,n}$
for system $A$
is given by a unitary operation $\hat{U}$ acting on  system 
$A$ and an ancillary system $E$, which is initially prepared in 
a state $\ket{\Sigma}_E$, followed by a projection measurement 
on $E$ to determine whether the transformation is successful or not.
Hence the machine
$\{\ket{\Phi_i}_A\stackrel{\gamma_i}{\longrightarrow}\ket{\Psi_i}_A\}_{i=1\ldots,n}$
exists iff there are a unitary $\hat{U}$, 
normalized states $\{\ket{P^{(i)}}_E\}$, and 
unnormalized states $\{\ket{\Omega_i}_{AE}\}$ such that 
\begin{equation}
\hat{U}(\ket{\Phi_i}_A \ket{\Sigma}_E)=
\sqrt{\gamma_i} \ket{\Psi_i}_A \ket{P^{(i)}}_E
+\ket{\Omega_i}_{AE}
\label{eq:th1proof1}
\end{equation}
and 
\begin{equation}
{}_E\bra{P^{(j)}} \ket{\Omega_i}_{AE} = 0
\label{eq:th1proof2}
\end{equation}
for all $i, j$.
Taking the inner products between the equations 
(\ref{eq:th1proof1})
with 
different values of $i$, we obtain
\begin{equation}
X=\sqrt{\Gamma} Y \sqrt{\Gamma} + \Omega,
\end{equation}
where matrix $\Omega:=[\braket{\Omega_i}{\Omega_j}]$ is 
positive semidefinite \cite{DG98}. Hence it is necessary that 
$X- \sqrt{\Gamma} Y \sqrt{\Gamma}$ be positive semidefinite.
Conversely, if $X- \sqrt{\Gamma} Y \sqrt{\Gamma}$ is 
positive semidefinite for a given set of $\{\ket{P^{(i)}}_E\}$,
%it was shown \cite{DG98} that 
there exist 
a unitary $\hat{U}$ and 
unnormalized states $\{\ket{\Omega_i}_{AE}\}$ satisfying 
Eqs.~(\ref{eq:th1proof1}) and (\ref{eq:th1proof2}),
as shown in Ref.~\cite{DG98}.
Theorem 1 is thus proved.

When the initial state is chosen from the set $\{\ket{\Phi_i}\}$
with 
{\it a priori} probability $p_i$, we may define the overall success probability 
$\gamma_{\rm tot}$
of a machine as 
\begin{equation}
\gamma_{\rm tot}:= \sum_i p_i \gamma_i.
\end{equation}
In this case, we can define the maximum 
success probability $\gamma_{\rm totmax}$
as 
\begin{equation}
 \gamma_{\rm totmax}:=\max_{\{\gamma_i\}}  \sum_i p_i \gamma_i,
\end{equation}
where the maximum is taken over all combinations $\{\gamma_i\}$
for which there exists a machine
$\{\ket{\Phi_i}\stackrel{\gamma_i}{\longrightarrow}\ket{\Psi_i}\}_{i=1,\ldots,n}$.

When the number of possible input states is 2, we can 
explicitly determine the achievable 
region $(\gamma_1,\gamma_2)$ from theorem 1.

{\it Corollary 1.} 
Let $\eta_{\rm in}:=|\braket{\Phi_1}{\Phi_2}|$ and 
$\eta_{\rm out}:=|\braket{\Psi_1}{\Psi_2}|$.
There exists a machine 
$\{\ket{\Phi_i}\stackrel{\gamma_i}{\longrightarrow}\ket{\Psi_i}\}_{i=1,2}$
if and only if $\gamma_1\ge 0$, $\gamma_2\ge 0$, and 
\begin{equation}
 \sqrt{(1-\gamma_1)(1-\gamma_2)}-\eta_{\rm in}
+\eta_{\rm out}\sqrt{\gamma_1\gamma_2}\ge 0.
\label{general2PT}
\end{equation}

{\em Proof.} Without loss of generality, we may
assume $\eta_{\rm in}=\braket{\Phi_1}{\Phi_2}$ and 
$\eta_{\rm out}=\braket{\Psi_1}{\Psi_2}$.
Since ${\rm Tr}[X-\sqrt{\Gamma}Y\sqrt{\Gamma}]\ge 0$,
$X-\sqrt{\Gamma}Y\sqrt{\Gamma}$ is positive semidefinite
iff $\det[X-\sqrt{\Gamma}Y\sqrt{\Gamma}]\ge 0$,
or equivalently,
\begin{eqnarray}
\sqrt{(1-\gamma_1)(1-\gamma_2)} \nonumber 
-\left|\eta_{\rm in}- \eta_{\rm out}
\braket{P^{(1)}}{P^{(2)}}\sqrt{\gamma_1\gamma_2}\right| 
\ge 0.
\nonumber \\
\label{eq:det1}
\end{eqnarray} 
Since the left-hand side (LHS) of Eq.~(\ref{eq:det1}) is no 
larger than the LHS of Eq.~(\ref{general2PT}), 
Eq.~(\ref{general2PT}) is necessary for the existence of a machine.
Conversely, 
whenever Eq.~(\ref{general2PT}) holds
and $\eta_{\rm in}\ge\eta_{\rm out}\sqrt{\gamma_1\gamma_2}$,
we can satisfy Eq.~(\ref{eq:det1}) by 
choosing $\braket{P^{(i)}}{P^{(j)}}=1$ and 
there exists a machine. 
When $\eta_{\rm in} < \eta_{\rm out}\sqrt{\gamma_1\gamma_2}$,
we can satisfy Eq.~(\ref{eq:det1}) by choosing
$\braket{P^{(1)}}{P^{(2)}}=\eta_{\rm in}/
(\eta_{\rm out}\sqrt{\gamma_1\gamma_2})$,
and hence there exists a machine also in this case,
proving the corollary.

When $\eta_{\rm in} > \eta_{\rm out}$, 
the region $(\gamma_1, \gamma_2)$ determined by Eq.~(\ref{general2PT}) is convex, and is bounded by 
the line $\gamma_1=0$, the line $\gamma_2=0$, and 
the curve specified by the equality in Eq.~(\ref{general2PT}),
which connects the points $(\gamma_1,\gamma_2)=(0,1-\eta_{\rm in}^2)$
and $(\gamma_1,\gamma_2)=(1-\eta_{\rm in}^2,0)$
through the point $\gamma_1=\gamma_2=(1-\eta_{\rm in})/(1-\eta_{\rm out})$.
When $\eta_{\rm in} \le \eta_{\rm out}$, $\gamma_1=\gamma_2=1$
satisfies Eq.~(\ref{general2PT}), namely, 
a deterministic machine 
$\{\ket{\Phi_i}\stackrel{1}{\longrightarrow}\ket{\Psi_i}\}_{i=1,2}$
exists. Note that Eq.~(\ref{general2PT}) still forbids regions 
of $(\gamma_1,\gamma_2)$ close to $(1,0)$ and $(0,1)$, reflecting 
the indistinguishability of the two input states.

\section{Probabilistic cloning of two  states
with supplementary information}\label{se:3}
%As we mentioned above, when the number of the input states is two, the relation
%\begin{equation}
%\mbox{scenario (I)} = \mbox{scenario (II)} \neq \mbox{scenario (III)}
%\end{equation}
%holds. 
In this section, we consider the case where
one makes $m$ copies of states
$\{ \ket{\psi_1},\ket{\psi_2} \}$ with the help of 
supplementary information in the form of pure states 
$\{ \ket{\phi_1},\ket{\phi_2} \}$.
We show that it is always better to try first 
the production of 
$m-1$ copies of the original information from the supplementary information
alone, independently of the original state,
which is implied by the following 
theorem. 
%\textit{{\bf Theorem 2:} Let the original information $\{ \ket{\psi_1},\ket{\psi_2} \}$  be a set of two different pure states. Let the supplementary information $\{ \ket{\phi_1},\ket{\phi_2} \}$ be an other set of two pure states satisfying $| \braket{\phi_1}{\phi_2} | \ge  | \braket{\psi_1}{\psi_2} |^{m-1}$. Then, scenario (I) is equivalent to scenario (II). }\\
%\textit{{\bf Theorem 2:} Let the original information $\{ \ket{\psi_1},\ket{\psi_2} \}$  be a set of two different pure states. Let the supplementary information $\{ \ket{\phi_1},\ket{\phi_2} \}$ be an other set of two pure states. Then, the success probability of scenario (I) is equivalent to that of scenario (II) for arbitrary prior probability $\{p_1,p_2\}$. }\\

{\it Theorem 2.}
%Suppose that $|\braket{\phi_1}{\phi_2}|>
%|\braket{\psi_1}{\psi_2}|^{m-1}$.
If there exists a machine 
$$
\{\ket{\psi_i}\ket{\phi_i}\stackrel{\gamma_i}{\longrightarrow}
\ket{\psi_i}^{\otimes m}\}_{i=1,2},
$$
then there exist a machine 
$$
\{\ket{\psi_i}\stackrel{\gamma_i^A}{\longrightarrow}
\ket{\psi_i}^{\otimes m}\}_{i=1,2}
$$
and a machine
$$
\{\ket{\phi_i}\stackrel{\gamma_i^B}{\longrightarrow}
\ket{\psi_i}^{\otimes m-1}\}_{i=1,2}
$$
with
\begin{equation}
 \gamma_i^B+ (1-\gamma_i^B)\gamma_i^A
\ge \gamma_i
 \;\; (i=1,2).
\end{equation}

%Note that we have excluded the uninteresting cases with 
%$|\braket{\phi_1}{\phi_2}|\le
%|\braket{\psi_1}{\psi_2}|^{m-1}$, in which $m-1$ copies
%can be deterministically produced from the supplementary 
% state $\ket{\phi_i}$ alone.

Before the proof of this theorem, several remarks may be in order.
If the original information is held by
Alice, and the supplementary information by Bob, 
theorem 2 implies that the optimal performance 
is always achieved just by one-bit classical communication
from Bob to Alice as follows: Bob, who possesses the supplementary state 
$\ket{\phi_i}$, first runs the machine 
$\{\ket{\phi_i}\stackrel{\gamma_i^B}{\longrightarrow}
\ket{\psi_i}^{\otimes m-1}\}_{i=1,2}$, and tells Alice 
whether the trial was successful or not. 
In the successful case, Alice just leaves her state $\ket{\psi_i}$
as it is, and hence they obtain $m$ copies in total.
If Bob's attempt has failed, Alice runs the machine 
$\{\ket{\psi_i}\stackrel{\gamma_i^A}{\longrightarrow}
\ket{\psi_i}^{\otimes m}\}_{i=1,2}$.
The total success probability for input state $\ket{\psi_i}\ket{\phi_i}$ 
in this protocol is given by
$\gamma_i^B+ (1-\gamma_i^B)\gamma_i^A$. 
Hence, by theorem 2, we see that the above protocol 
is as good as any other protocol in which Alice 
and Bob communicate through quantum channels.
Note that when $\{ \ket{\psi_i} \}$ includes no
pair of identical states, $\lim_{m\to\infty}
\braket{\psi_i}{\psi_j}^m=\delta_{ij}$ holds for
any $i \neq j$. 
Hence in the limit $m\to\infty$
the machine $\{\ket{\psi_i}\ket{\phi_i}
\stackrel{\gamma_i}{\longrightarrow}\ket{\psi_i}^{\otimes
m}\}_{i=1,\ldots,n}$ effectively
carries out unambiguous discrimination of the set
$\{\ket{\psi_i}\ket{\phi_i} \}$. 
Therefore, in this limit theorem 2 reproduces the results in Ref.~\cite{CY01},
namely, local operations and classical communication achieves the global optimality of unambiguous discrimination of any two pure product states with arbitrary {\it a priori} probability $p_i$.

When the initial state $\ket{\psi_i}\ket{\phi_i}$  is chosen 
with probability $p_i$, it follows from theorem 2 that the maximum 
overall success probability $\gamma_{\rm totmax}$ is achieved by the above two-step protocol.
For a special case of 
$p_1=p_2=1/2$, we can directly confirm this as follows.
The maximum overall success probability 
$\gamma_{\rm totmax}$
can easily be calculated 
by optimizing $(\gamma_1+\gamma_2)/2$ over the region 
in corollary 1, and it is found to be 
\begin{equation}
\gamma_{\rm totmax}=\frac{1-|\alpha\beta|}{1-|\alpha|^m},
\end{equation}
where $\alpha:=\braket{\psi_1}{\psi_2}$ and 
$\beta:=\braket{\phi_1}{\phi_2}$.
Corollary 1 also 
shows the existence of a machine 
$\{\ket{\phi_i}\stackrel{\gamma_i^B}{\longrightarrow}
\ket{\psi_i}^{\otimes m-1}\}_{i=1,2}$
with 
\begin{equation}
 \gamma_1^B=\gamma_2^B=\frac{1-|\beta|}{1-|\alpha|^{m-1}}
\end{equation}
and a machine $\{\ket{\psi_i}\stackrel{\gamma_i^A}{\longrightarrow}
\ket{\psi_i}^{\otimes m}\}_{i=1,2}$ with 
\begin{equation}
 \gamma_1^A=\gamma_2^A=\frac{1-|\alpha|}{1-|\alpha|^{m}}.
\end{equation}
Hence, using these machines in the two-step protocol,
we obtain an overall 
success probability 
\begin{equation}
 \gamma_1^B+ (1-\gamma_1^B)\gamma_1^A=\frac{1-|\alpha\beta|}{1-|\alpha|^m},
\end{equation}
which coincides with $\gamma_{\rm totmax}$.
For cases with general $(p_1,p_2)$, it is even difficult to 
represent $\gamma_{\rm totmax}$ in an explicit form,
but theorem 2 states that $\gamma_{\rm totmax}$ is
always achieved by the two-step protocol.

{\em Proof of Theorem 2.} 
When $|\braket{\phi_1}{\phi_2}|\le |\braket{\psi_1}{\psi_2}|^{m-1}$, 
from corollary 1, there exists a machine 
$\{\ket{\phi_i}\stackrel{\gamma_i^B}{\longrightarrow}
\ket{\psi_i}^{\otimes m-1}\}_{i=1,2}$
with $\gamma_1^B=\gamma_2^B=1$, and theorem 2 obviously holds.
We thus assume $|\braket{\phi_1}{\phi_2}|>
|\braket{\psi_1}{\psi_2}|^{m-1}$ in the following.

Let ${\cal R}$ be the region of points $(\gamma_1, \gamma_2)$
for which a machine 
$\{\ket{\psi_i}\ket{\phi_i}\stackrel{\gamma_i}{\longrightarrow}
\ket{\psi_i}^{\otimes m}\}_{i=1,2}$ exists.
We first show that it suffices to prove theorem 2 for the cases where
$\gamma_1\ge \gamma_2$ and 
$(\gamma_1,\gamma_2)$ is on a boundary of 
the achievable region 
${\cal R}$, namely (see corollary 1),
\begin{equation}
 \sqrt{(1-\gamma_1)(1-\gamma_2)}-|\alpha\beta|
+|\alpha|^m\sqrt{\gamma_1\gamma_2}= 0.
 \label{eq:2.1}
\end{equation}
For any other point $(\gamma'_1, \gamma'_2)$
in the region ${\cal R}$ with 
$\gamma'_1\ge \gamma'_2$, we can find a point
$(\gamma_1, \gamma_2)$ on the boundary with $\gamma_1\ge \gamma_2$
satisfying 
$\gamma'_i=y\gamma_i \; (i=1,2)$ with $y\le 1$.
If theorem 2 holds for $(\gamma_1, \gamma_2)$,
there are machines with $\gamma_i^A$ and $\gamma_i^B$
satisfying $\gamma_i^B+(1-\gamma_i^B)\gamma_i^A\ge \gamma_i\ge \gamma'_i$.
This implies that theorem 2 also holds for $(\gamma'_1, \gamma'_2)$.
The cases $\gamma'_1< \gamma'_2$ follow from the symmetry.

Consider a point $(\gamma_1, \gamma_2)$ on the boundary and 
satisfying $\gamma_1\ge\gamma_2$.
Let $x:=\gamma_2/\gamma_1$.
From Eq.~(\ref{eq:2.1}) we have
\begin{equation}
 \sqrt{(1-\gamma_1)(1-x\gamma_1)}-|\alpha\beta|
+|\alpha|^m\sqrt{x}\gamma_1= 0.
 \label{eq:2.2}
\end{equation}
For the machine  
$\{\ket{\phi_i}\stackrel{\gamma_i^B}{\longrightarrow}
\ket{\psi_i}^{\otimes m-1}\}_{i=1,2}$,
we choose $(\gamma_1^B,\gamma_2^B)$ as the point 
satisfying $\gamma_2^B=x\gamma_1^B$ and being on the 
boundary (for this machine), 
namely, satisfying 
\begin{multline}
  \sqrt{(1-\gamma_1^B)(1-\gamma_2^B)}-|\beta|
+|\alpha|^{m-1}\sqrt{\gamma_1^B\gamma_2^B}  \\
= \sqrt{(1-\gamma_1^B)(1-x \gamma_1^B)}-|\beta|
+|\alpha|^{m-1}\sqrt{x}\gamma_1^B
= 0.\label{eq:bb}
\end{multline}
Let us define $(\gamma_1^A,\gamma_2^A)$ by 
\begin{eqnarray}
 1-\gamma_1^A&:=&\frac{1-\gamma_1}{1-\gamma_1^B},
\label{eq:2.7}
\\
 1-\gamma_2^A&:=&\frac{1-\gamma_2}{1-\gamma_2^B}
=\frac{1-x \gamma_1}{1-x \gamma_1^B}.
\label{eq:2.8}
\end{eqnarray}
For this choice, 
$\gamma_i^B+(1-\gamma_i^B)\gamma_i^A= \gamma_i$ holds for $i=1,2$. 
Hence we only have to show the existence of 
a machine 
$\{\ket{\psi_i}\stackrel{\gamma_i^A}{\longrightarrow}
\ket{\psi_i}^{\otimes m}\}_{i=1,2}$ with 
$(\gamma_1^A,\gamma_2^A)$ defined above.

Now, consider a protocol in which Bob runs 
machine $\{\ket{\phi_i}\stackrel{\gamma_i^B}{\longrightarrow}
\ket{\psi_i}^{\otimes m-1}\}_{i=1,2}$ and 
Alice does nothing. This protocol can be viewed as 
a machine 
$\{\ket{\psi_i}\ket{\phi_i}\stackrel{\gamma_i^B}{\longrightarrow}
\ket{\psi_i}^{\otimes m}\}_{i=1,2}$,
and hence $(\gamma_1^B,\gamma_2^B)$ is in the region
${\cal R}$. Then, the points $(0,0)$, 
$(\gamma_1^B,\gamma_2^B)$, $(\gamma_1,\gamma_2)$
should be on a straight line in this order,
and $\gamma_1^B \le \gamma_1$ holds.
Hence, from Eqs.~(\ref{eq:2.7}) and (\ref{eq:2.8}), we have
\begin{eqnarray}
 \gamma_1^A\ge 0, \;\; \gamma_2^A\ge 0.
\end{eqnarray}

Using Eqs. (\ref{eq:2.2})--(\ref{eq:2.8}), we obtain
\begin{eqnarray}
\sqrt{(1-\gamma_1^A)(1-\gamma_2^A)}
&=&\frac{\sqrt{(1-\gamma_1)(1-x\gamma_1)}}
{\sqrt{(1-\gamma_1^B)(1-x\gamma_1^B)}}
\nonumber \\
&=&|\alpha|\frac{|\beta|-|\alpha|^{m-1}\sqrt{x}\gamma_1}
{|\beta|-|\alpha|^{m-1}\sqrt{x}\gamma_1^B}
\end{eqnarray}
and 
\begin{eqnarray}
 \sqrt{\gamma_1^A\gamma_2^A}
&=&\frac{\sqrt{x}(\gamma_1-\gamma_1^B)}
{\sqrt{(1-\gamma_1^B)(1-x\gamma_1^B)}}
\nonumber \\
&=&\frac{\sqrt{x}(\gamma_1-\gamma_1^B)}
{|\beta|-|\alpha|^{m-1}\sqrt{x}\gamma_1^B}.
\end{eqnarray}
Hence it is not difficult to show that
\begin{equation}
  \sqrt{(1-\gamma_1^A)(1-\gamma_2^A)}-|\alpha|
+|\alpha|^{m}\sqrt{\gamma_1^A\gamma_2^A}=0.
\end{equation}
From corollary 1, there exists a machine 
$\{\ket{\psi_i}\stackrel{\gamma_i^A}{\longrightarrow}
\ket{\psi_i}^{\otimes m}\}_{i=1,2}$ with 
$(\gamma_1^A,\gamma_2^A)$, and theorem 2 is proved.

%The theorem  can be interpreted as that the probabilistic cloning has no ``non%-locality'' when the number of the input 
%states is two.
%Moreover, it is worth noting that, from Eq.~(\ref{eq:2.1.1}), the smaller the %inter-product $|\beta|=|\braket{\phi_1}{\phi_2}|$ is, the bigger the maximum e%fficiency $\Gamma^{{\footnotesize (I)}}_{\mbox{{\footnotesize max}}}$ is. It i%plies that the supplementary information $\ket{\phi_i}$ is useful for probabilistic cloning in contrast to deterministic cloning. 

\section{Probabilistic cloning with supplementary information for three
 or more states}\label{se:4}

When the number of the possible states is 3 or more,
Theorem 2 is not always true, and there may exist a better 
protocol than just running machines  
$\{\ket{\phi_i}\stackrel{\gamma_i^B}{\longrightarrow}
\ket{\psi_i}^{\otimes m-1}\}_{i=1,2,\ldots,n}$
and 
$\{\ket{\psi_i}\stackrel{\gamma_i^A}{\longrightarrow}
\ket{\psi_i}^{\otimes m}\}_{i=1,2,\ldots,n}$.
We will give such an example in this section, and 
also show that a somewhat stronger statement holds
about how the supplementary and the original information 
should be combined to give the optimal performance.
For this purpose, we assume that two separated parties, 
Alice and Bob, have the original information $\ket{\psi_i}$ 
and the supplementary information $\ket{\phi_i}$, respectively.
We do not care which of the parties produces the copies,
 as long as they produce $m$ copies of $\ket{\psi_i}$ in total,
namely, the task is successful when 
\begin{equation}
 \ket{\psi_i}_A \ket{\phi_i}_B \longrightarrow \ket{\psi_i}^{\otimes m-k}_A \ket{\psi_i}^{\otimes k}_B, \\ (i=1,2,\ldots,n),
\end{equation}
for any integer $k$.
We consider two scenarios depending on the allowed communication 
between Alice and Bob.

{\it Scenario I.} Alice and Bob can use a one-way quantum channel from Bob to Alice.
Note that this scenario is equivalent to the case where a single party having both the original and the supplementary information runs a machine
$\{\ket{\psi_i}\ket{\phi_i}\stackrel{\gamma_i}{\longrightarrow}\ket{\psi_i}^{\otimes
m}\}_{i=1,\ldots,n}$,
and its success probabilities are determined by theorem 1.

{\it Scenario II.} Alice and Bob can use only a one-way classical channel from Bob to Alice.
Note that the two-step protocol in the last section is 
included in this scenario.
In what follows, we construct 
an example showing a gap between the two scenarios.

Consider an $n$-dimensional Hilbert space, and choose an 
orthonormal basis $\{\ket{j}\}_{j=1,\ldots,n}$.
Let us define $n$ normalized states $\{\ket{\mu_j}\}_{j=1,\ldots,n}$
as follows:
\begin{equation}
 \ket{\mu_j}:= \sqrt{1-(n-1)z^2}\ket{j}-z\sum_{i\neq j}\ket{i},
\end{equation}
where $z\ge0$. The inner product between any pair of the states is given by
\begin{equation}
 \braket{\mu_i}{\mu_j}=z\left[(n-2)z-2\sqrt{1-(n-1)z^2}\right]
\end{equation}
for $i \neq j$.
The right-hand side is zero for  $z=0$, and 
is $-1/(n-1)$ for $z=1/\sqrt{n(n-1)}$. 
By continuity, we see that for any $\alpha\in [-1/(n-1),0]$,
there exists a set of $n$ normalized states $\{\ket{\mu_j}\}_{j=1,\ldots,n}$
satisfying $\braket{\mu_i}{\mu_j}=\alpha$ for $i\neq j$.

Now we consider a problem of producing $m$ copies of a state
chosen randomly ($p_j=1/n$) from the set 
$\{\ket{\psi_j}\}_{j=1,\ldots,n}$ satisfying 
\begin{equation}
 \braket{\psi_i}{\psi_j}=\alpha=-|\alpha| \;\; \left(0<|\alpha|<\frac{1}{n-1}\right)
\label{psi-ind}
\end{equation}
for any $i\neq j$, each accompanied by supplementary 
information $\{\ket{\phi_j}\}_{j=1,\ldots,n}$ satisfying 
\begin{equation}
 \braket{\phi_i}{\phi_j}=\beta=-\frac{1}{n-1}
\label{phi-dep}
\end{equation}
for any $i\neq j$. 
Both sets of states, $\{\ket{\psi_j}\}_{j=1,\ldots,n}$ and
$\{\ket{\phi_j}\}_{j=1,\ldots,n}$,
exist because they are special cases of the set
$\{\ket{\mu_j}\}_{j=1,\ldots,n}$ above.

Let $\gamma_{\rm totmax}^{I}$ and $\gamma_{\rm totmax}^{II}$
be the maximum overall probabilities in scenarios I and II, respectively. We show that for any $n\ge 3$
and any $m\ge 2$, there is a gap between the two scenarios
$(\gamma_{\rm totmax}^{I}>\gamma_{\rm totmax}^{II})$ for 
sufficiently small (but nonzero) $|\alpha|$.

First, we derive a lower bound for $\gamma_{\rm totmax}^{I}$, 
written as 
\begin{equation}
 \gamma_{\rm totmax}^{I} \ge \frac{1-|\beta||\alpha|}
{1-|\beta||\alpha|^m}=\frac{n-1-|\alpha|}{n-1-|\alpha|^m}.
\label{lowerbound}
\end{equation}
This relation can be proved via theorem 1 with $\ket{\Phi_i}=\ket{\psi_i}\ket{\phi_i}$ and $\ket{\Psi_i}=\ket{\psi_i}^{\otimes m}$ as follows.
We separate the cases depending on the parity of $m$.

When $m$ is even, we take 
\begin{equation}
 \gamma_i = \frac{1-|\beta||\alpha|}
{1-|\alpha|^m} \left( > \frac{1-|\beta||\alpha|}
{1-|\beta||\alpha|^m} \right)
\label{meven}
\end{equation}
and $\braket{P_i}{P_j}=1$ for any $i,j$.
Then we obtain 
\begin{equation}
X- \sqrt{\Gamma} Y\sqrt{\Gamma}=\frac{|\beta||\alpha|-|\alpha|^m}{1-|\alpha|^m} Z, \label{eq:matI}
\end{equation}
where $Z$ is an $n\times n$ matrix defined by
\begin{equation}
Z:=\left(
  \begin{array}{ccc}
    1   &  \cdots   & 1   \\
    \vdots &  \ddots   &   \vdots \\
    1   &  \cdots  & 1   \\
  \end{array}
\right).
\end{equation}
Since the eigenvalues of $Z$ are only $0$ ($n-1$ degeneracy) and $n$ (no degeneracy),
$Z$ is positive semidefinite and so is $X- \sqrt{\Gamma} Y\sqrt{\Gamma}$.
Hence, by theorem 1, there
exists a machine satisfying Eq.~(\ref{meven}), and 
Eq.~(\ref{lowerbound}) holds in this case.

When $m$ is odd, we take 
\begin{equation}
 \gamma_i = \frac{1-|\beta||\alpha|}
{1-|\beta||\alpha|^m}
\label{modd}
\end{equation}
and $\braket{P_i}{P_j}=\beta=-1/(n-1)$ for any $i\neq j$.
Then we obtain 
\begin{equation}
X- \sqrt{\Gamma} Y\sqrt{\Gamma}=\frac{1-|\alpha|^{m-1}}{1-|\beta||\alpha|^m} |\beta| |\alpha| Z,
\end{equation}
which is positive semidefinite. By theorem 1, there
exists a machine satisfying Eq.~(\ref{modd}), and 
Eq.~(\ref{lowerbound}) holds also in this case,
namely, irrespective of $m$.

Next, we derive an upper bound on $\gamma_{\rm totmax}^{II}$.
We start by proving the following lemma.

{\it Lemma 1.} Consider a linearly independent set of $n$ states
$\{\ket{\Psi_1},\ket{\Psi_2},\ldots,\ket{\Psi_n}\}$, and another set of 
$n$ states $\{\ket{\Phi_1},\ket{\Phi_2},\ldots,\ket{\Phi_n}\}$
satisfying
\begin{equation}
\sum_{i=1}^n b_i \ket{\Phi_i}=0.
\label{eq:3.1}
\end{equation}
If $b_j \neq 0$, there is no machine 
$\{\ket{\Phi_i}\stackrel{\gamma_i}{\longrightarrow}
\ket{\Psi_i}\}_{i=1,\ldots, n}$
with $\gamma_j>0$.

{\em Proof.} Suppose that there exists a machine
$\{\ket{\Phi_i}\stackrel{\gamma_i}{\longrightarrow}
\ket{\Psi_i}\}_{i=1,\ldots, n}$. 
From theorem 1,
$n \times n$ matrix 
$X- \sqrt{\Gamma} Y \sqrt{\Gamma}$ is positive semidefinite.
Then, for the vector $\bm{b}=(b_1,b_2,\ldots,b_n)^{T}$ 
satisfying the Eq.~(\ref{eq:3.1}), we have 
\begin{equation}
\bm{b}^{\dag}(X- \sqrt{\Gamma} Y \sqrt{\Gamma})
\bm{b} \ge 0.
\end{equation}
Since $\bm{b}^{\dag} X \bm{b}=0$ from
Eq.~(\ref{eq:3.1}), we have 
$\bm{b}^{\dag} (\sqrt{\Gamma} Y \sqrt{\Gamma}) \bm{b}\le 0$.
Since the linear independence of $\{\ket{\Psi_i}\}$ implies that 
$Y$ is positive definite, it follows that $\sqrt{\Gamma}
\bm{b}=\bm{0}$, and hence $b_i\gamma_i=0$ for all $i$.
Then, $b_j \neq 0$ implies $\gamma_j=0$. 

In the problem at hand, 
the set of states $\{\ket{\psi_i}^{\otimes k}\}$ 
is linearly independent for any integer $k\ge 1$ since 
the eigenvalues of 
the $n\times n$ matrix
$[\braket{\psi_i}{\psi_j}^{k}]=(1-\alpha^k)I+\alpha^k Z$ are
only $1-\alpha^{k}>0$ ($n-1$ degeneracy) 
and $1+(n-1)\alpha^{k}>0$ (no degeneracy),
where $I$ 
is the 
$n \times n$ identity matrix.
The set $\{\ket{\phi_i}\}$ satisfies 
$\sum_i \ket{\phi_i}=0$ 
since $\sum_{i,j}\braket{\phi_i}{\phi_j}=n+n(n-1)\beta=0$.
Then, we see from lemma 1 that any machine
$\{\ket{\phi_i}\stackrel{\gamma^B_i}{\longrightarrow}
\ket{\psi_i}^{\otimes k}\}_{i=1,\ldots, n}$
has zero success probability, 
$\gamma^B_1=\gamma^B_2=\cdots=\gamma^B_n=0$.
In scenario II, this fact implies that 
all of the $m$ copies must be produced by Alice, 
and Bob's role is just to provide classical information
to help Alice's operation. Hence, we are allowed to 
limit Bob's action to a POVM measurement $\{\hat{E_\mu}\}$
applied to his initial state $\ket{\phi_i}$, providing
outcome $\mu$ with probability $p_{\mu|i}:=\bra{\phi_i} 
\hat{E_\mu}\ket{\phi_i}$.
Depending on the outcome $\mu$ received from Bob,
Alice runs a machine 
$\{\ket{\psi_i}\stackrel{\gamma^{(\mu)}_{i}}{\longrightarrow}
\ket{\psi_i}^{\otimes m}\}_{i=1,\ldots, n}$
to produce $m$ copies of state $\ket{\psi_i}$.
Since the initial state $\ket{\psi_i}$ is randomly chosen
($p_i=1/n$), 
the overall success probability is 
\begin{equation}
 \gamma^{II}_{\rm tot}=\sum_\mu p_\mu 
\left(
\sum_{i=1}^{n}p_{i|\mu}\gamma^{(\mu)}_{i} 
\right),
\end{equation}
where $p_\mu:=\sum_i p_{\mu|i}p_i$ and $p_{i|\mu}:=p_{\mu|i}p_i/p_{\mu}$.

From theorem 1, $\Gamma^{(\mu)}:=\mbox{diag}
(\gamma^{(\mu)}_{1},\gamma^{(\mu)}_{2},\ldots,\gamma^{(\mu)}_{n})$
should satisfy 
\begin{equation}
\bm{b}^{\dag}\left(X- \sqrt{\Gamma^{(\mu)}} Y \sqrt{\Gamma^{(\mu)}}\right)
\bm{b} \ge 0
\end{equation}
for any $\bm{b}$. Here the elements of matrices $X$ and $Y$ are 
given by $X_{ij}=(1-\alpha)\delta_{i,j}+\alpha$
and $Y_{ij}=[(1-\alpha^m)\delta_{i,j}+\alpha^m]
\braket{P^{I}}{P^{(j)}}$.
If we choose
$\bm{b}=(\sqrt{p_{1|\mu}},\sqrt{p_{2|\mu}},\ldots,\sqrt{p_{n|\mu}})$, we have
\begin{eqnarray}
0 &\le& 
 1-\sum_i p_{i|\mu} \gamma_i^{(\mu)}
 - |\alpha| \sum_{i,j(\neq i)} \sqrt{p_{i|\mu} p_{j|\mu}}\nonumber\\
&&- \alpha^m \sum_{i,j(\neq i)}\sqrt{p_{i|\mu} p_{j|\mu}
 \gamma_i^{(\mu)} \gamma_j^{(\mu)}}\braket{P^{I}}{P^{(j)}} \nonumber\\
&\le&1-\sum_i p_{i|\mu} \gamma^{(\mu)}_i 
 - |\alpha| \sum_{i,j(\neq i)} \sqrt{p_{i|\mu} p_{j|\mu}}\nonumber\\
&&+ (n-1) |\alpha|^m \sum_i p_{i|\mu} \gamma^{(\mu)}_i, \label{eq:4.0.5}
\end{eqnarray}
where we have used $2 \sqrt{p_{i|\mu} p_{j|\mu}
 \gamma_i^{(\mu)} \gamma_j^{(\mu)}}\le p_{i|\mu}\gamma_i^{(\mu)} 
+ p_{j|\mu}\gamma_j^{(\mu)}$ and $|\braket{P^{I}}{P^{(j)}}|\le 1$.
Using this relation, we obtain
\begin{equation}
\gamma^{II}_{\rm totmax} \le \frac{1
-|\alpha| \sum_{i,j(\neq i)}
\sum_\mu p_\mu
 \sqrt{p_{i|\mu} p_{j|\mu}} }{1-(n-1)|\alpha|^m}. \label{eq:4.3}
\end{equation}
We further bound the term in the numerator by 
using 
$p_{i|\mu} p_{\mu}=p_{\mu|i} p_{i}=\bra{\phi_i} \hat{E_{\mu}}
\ket{\phi_i}/n$,
the completeness relation $\sum_{\mu} \hat{E_{\mu}} =\hat{1}$, and 
the Cauchy-Schwarz inequality
\begin{eqnarray}
\sum_{\mu}p_{\mu} \sqrt{p_{i|\mu}  p_{j|\mu} } 
&=&\frac{1}{n} \sum_{\mu} \sqrt{\bra{\phi_i}\hat{E_{\mu}}\ket{\phi_i}\bra{\phi_j}\hat{E_{\mu}}\ket{\phi_j}} \nonumber \\
%&&=\frac{1}{n} \sum_{\mu} \sqrt{\bra{\phi_i}\sqrt{\hat{E_{\mu}}}\sqrt{\hat{E_{\mu}}}\ket{\phi_i}\bra{\phi_j}\sqrt{\hat{E_{\mu}}}\sqrt{\hat{E_{\mu}}}\ket{\phi_j}} \nonumber \\
&\ge&\frac{1}{n}\sum_{\mu} |\bra{\phi_i}\hat{E_{\mu} }\ket{\phi_j}| \nonumber \\
&\ge&\frac{1}{n}|\braket{\phi_i}{\phi_j}| 
=\frac{1}{n(n-1)}. \label{eq:4.4}
\end{eqnarray}
From Eqs. (\ref{eq:4.3}) and (\ref{eq:4.4}), we obtain 
\begin{equation}
\gamma^{II}_{\rm totmax} \le \frac{1
-|\alpha| }{1-(n-1)|\alpha|^m}. 
\label{upperbound}
\end{equation}

Combining Eqs.~(\ref{lowerbound}) and (\ref{upperbound}),
we obtain
\begin{equation}
 \gamma^{I}_{\rm totmax}-\gamma^{II}_{\rm totmax}
\ge 
\frac{(n-2)|\alpha|(1+|\alpha|^m-n|\alpha|^{m-1})}{(n-1-|\alpha|^m)[1-(n-1)|\alpha|^m]},
\end{equation}
which shows that $\gamma^{I}_{\rm totmax}>\gamma^{II}_{\rm totmax}$
when $|\alpha|$ is a small enough positive number. 

\section{summary}\label{se:5}
In this paper, we have discussed probabilistic cloning 
of a mutually nonorthogonal set of pure states
$\{\ket{\psi_1},\ldots,\ket{\psi_n}\}$, with the 
help of supplementary information. 
It has turned out that the situation is quite different 
for $n=2$ and for other cases. When $n=2$, the role 
of the supplementary information is limited to just produce 
copies on its own, independently of the original state. 
This property is quite similar to the property in  
deterministic cloning, stated in the stronger no-cloning theorem.
For $n\ge 3$, such a simple property does not hold any longer.
We assumed that the original and the supplementary information are held by 
separated parties, and asked what kind of communication 
is required to achieve the optimal performance.
We have found examples in which the optimum
performance cannot be achieved even if we allow any amount 
of classical communication from the party with 
the supplementary information to the other. If we limit 
to the one-way communication scenarios, this result means that
a nonclassical interaction between the supplementary and
the original information helps to improve the performance.
On the other hand, if we allow the flow of information in the 
other direction, we are not sure the gap still exists. 
Analysis of such two-way protocols will be an interesting 
problem. The cases where the set $\{\ket{\psi_i}\}$ includes
a mutually orthogonal pair, or the cases where
supplementary information is provided as a mixed state 
are also worth investigating.\\\\

\section*{Acknowledgements}
We thank R. Namiki, S.K. Ozdemir, and T. Yamamoto for helpful discussions.
This work was supported by 21st Century COE Program by the Japan Society for the Promotion of Science and by a MEXT Grant-in-Aid for Young Scientists (B) No.~17740265.


\begin{thebibliography}{99}%{}"à'ÉŽQl•¶Œ£'Ì'"'ð''­
\bibitem{WZ82}
    W.K. Wootters and W.H. Zurek, Nature (London) {\bf 299}, 802 (1982).
\bibitem{Y86}
    H.P. Yuen, Phys. Lett. {\bf 113A}, 405 (1986).
\bibitem{BH96}
    V. Buzek and M. Hillery, Phys. Rev. A {\bf 54}, 1844 (1996).
\bibitem{MPH97}
    D. Mozyrsky, V. Privman, and M. Hillery, Phys. Lett. A {\bf 226}, 253 (1997).
\bibitem{GH97}
    N. Gisin and B. Huttner, Phys. Lett. A {\bf 228}, 13 (1997).
\bibitem{BVOKH97}
    V. Buzek, V. Vedral, M.B. Plenio, P.L. Knight, and M. Hillery, Phys. Rev. A {\bf 55}, 3327 (1997).
\bibitem{HB97}
    M. Hillery and V. Buzek, Phys. Rev. A {\bf 56}, 1212 (1997).
\bibitem{BDEMS98}
    D. Bru\ss, D.P. DiVincenzo, A. Ekert, C.A. Fuchs, C. Macchiavello, and J.A. Smolin, Phys. Rev. A {\bf 57}, 2368 (1998).
\bibitem{GM97}
    N. Gisin and S. Massar, Phys. Rev. Lett. {\bf 79}, 2153 (1997).
\bibitem{BEM98}
    D. Bru\ss, A. Ekert, and C. Macchiavello, Phys. Rev. Lett. {\bf 81}, 2598 (1998).
\bibitem{DG98}
    L.M. Duan and G.C. Guo, Phys. Rev. Lett. {\bf 80}, 4999 (1998).
\bibitem{J02}
    R. Jozsa, quant-ph/0204153.
%\bibitem{JCY05}
    %Z. Ji, H. Cao, and M. Ying, Phys. Rev. A {\bf 71}, 032323 (2005).
\bibitem{CY01}
    Y.X. Chen and D. Yang, Phys. Rev. A {\bf64}, 064303 (2001).  
\end{thebibliography}
\end{document}